\documentclass{article}
\usepackage{sao2,psfig}
\issue{2007, 62, 285-295}

\begin{document}
\markboth{Naselsky, Verkhodanov, Christensen, \& Chiang}
	 {On the Antenna Beam Shape Reconstruction Using Planet Transit}
\title{On the Antenna Beam Shape Reconstruction Using Planet Transit}
\author{ P.D. Naselsky\inst{a}
    \and O.V. Verkhodanov\inst{b}
    \and P.R. Christensen\inst{a}
    \and L.-Y. Chiang\inst{a}
}
\institute{
Niels Bohr Institute, Blegdamsvej 17, DK-2100 Copenhagen, Denmark
\and \saoname }
\date{December 18, 2006}{January 17, 2007}
\maketitle

\begin{abstract}
The calibration of the in-flight antenna beam shape and possible beam
degradation is one of the most crucial tasks for the upcoming
Planck mission. We examine several effects which could significantly
influence the in-flight main beam calibration using planet transit: the
problems of the variability of	the Jupiter's flux, the antenna temperature
and passing of the planets through the main beam. We estimate these
effects on the antenna beam shape calibration and calculate the limits
on the main beam and far sidelobe measurements,
using observations of Jupiter and
Saturn. We also discuss possible effects of degradation of the mirror
surfaces and specify corresponding parameters which can help us to
determine these effects.
\end{abstract}

\section{INTRODUCTION}

The ESA Planck Surveyor
\footnote{\texttt{http://astro.estec.esa.nl/SA-general/Projects/
Planck/}} is designed to image the whole sky of the Cosmic
Microwave Background (CMB) radiation with unprecedented sensitivity
($\Delta T/T \sim 2 \times 10^{-6}$) and angular resolution
(down to $5\arcmin$) at 9 frequencies:
30, 44, 70, 100, 143, 217, 353, 545, 857\,GHz at Low Frequency
Instrument (LFI) \cite{mandolesi:Verkhodanov_n}
and High Frequency Instrument (HFI)
\cite{puget:Verkhodanov_n}.

To achieve these high sensitivity and resolution, it is necessary to
carefully account for all potential systematic features in the data
\cite{bl:Verkhodanov_n}.

One of the systematic effects is related to the in-flight antenna beam
shape and its reconstruction. Apart from the need to
acquire the radiation pattern of the antenna beam before the flight, the
calibration of the in-flight antenna beam shape is one of the key
components
for achieving the scientific goals of the Planck mission. This
problem has been considered by different Planck groups
\cite{demaagt:Verkhodanov_n,burigana1997:Verkhodanov_n,burigana2001:Verkhodanov_n,beam:Verkhodanov_n},
and recently by
\cite{nas1:Verkhodanov_n,nas2:Verkhodanov_n}.

The accuracy of the CMB anisotropy $C_\ell$ estimation will be
affected, among other experiment parameters, by our ignorance of the
in-flight antenna beam shape of the main beam and far sidelobes, and
possible
degradation of the mirror surface shapes. When scanning
the antenna beam  of the Planck
mission is to move across the  sky, implying that the antenna
beam response is a function of time. After pixelization of the
time-ordered
data the position of each pixel on the pixelized CMB map is directly
related to some data points in the time stream. It is therefore
necessary to obtain the information about the in-flight beam shape,
its inclination and the location of the beam center relative to each
pixel, in order to improve the model of the in-flight main beam shape
as well as in the far sidelobe region.

To tackle this issue, Burigana et al. \cite{burigana2001:Verkhodanov_n}
have suggested
a method which
uses
planet transit to reconstruct the in-flight beam shape. These
planet crossings can help the in-flight beam recovery down to
$-$25$\div$32.5\,dB at 30\,GHz. They also
showed that the main beam
pattern can be described by the bivariate Gaussian
approximation. Recently Chiang et al. \cite{beam:Verkhodanov_n} have proposed
 another method for the
beam shape estimation based on the interplay of amplitudes and phases
of the sky signal and instrumental noise. This method is useful in
extraction of the antenna main beam shape down to $-$7$\div$10\,dB,
and does not need a strong radio source shape calibration. These
methods have laid a base for the determination of both the in-flight
antenna beam shape and its variations during observations.

The aim of this paper is to re-examine in details the proposed method
of the in-flight antenna shape reconstruction
by planet crossing \cite{burigana2001:Verkhodanov_n} in order to
estimate possible beam degradation effects. The method, based on
Jupiter and Saturn observations, has some subtleties needed to
be addressed such as the temporal and frequency variations of the Jupiter's
flux and passing of the planets through the main beam.

This paper is arranged as
follows. In Section 2 we describe the Planck scan strategy in
relation to planet observations, and the beam definitions. In Section 3
we discuss the in-flight beam calibration using planet transit. We
concentrate in Section 4 on the effects in beam calibration due to
fluctuation of the Jupiter's flux density. In Section 5 we discuss
variations arising from the distance to the planets and from the scan
strategy. Conclusion is in Section 6.

\section{SCAN STRATEGY AND THE BEAM PROBLEM}
The proposed scan strategy \cite{demaagt:Verkhodanov_n}
for the Planck mission is a
whole-sky scanning with the satellite orbiting around the L2
Lagrangian point of the Earth--Sun system. The satellite spin axis will
be
pointed in the anti-Sun direction and will have a tilt to the
ecliptic plane of 10\degr. The telescope optical angle is inclined
at $85\degr$ to the spin axis. The telescope will scan the same
circle 60 times around the spin axis at 1 r.p.m. Each hour the
spin axis will be shifted  along the ecliptic plane with
a step of 2\farcm5 during
the entire mission (about 15 months).

\begin{figure}[t!]
\centerline{
\psfig{figure=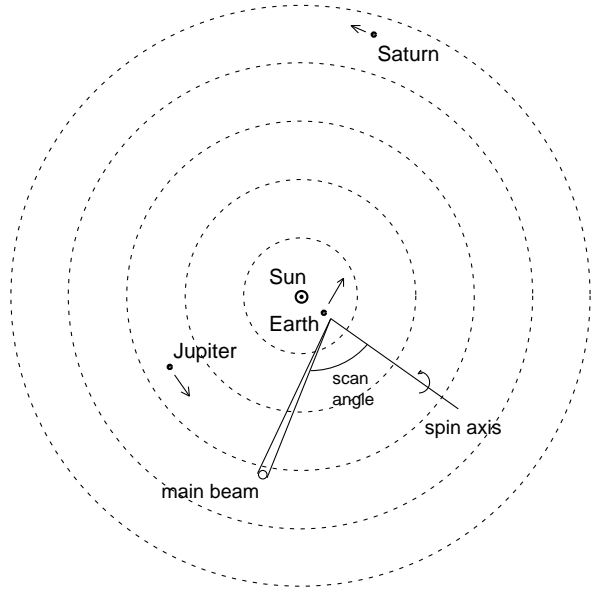,width=8cm,bbllx=50pt,bblly=270pt,%
bburx=550pt,bbury=760pt,clip=}
}
\caption{Positions of Jupiter and Saturn at the beginning of the Planck
mission (November, 2008).
The radius of each concentric circle is differed by 2~AU.}
\label{xephem:Verkhodanov_n}
\end{figure}

\subsection{ Observations of the Planets}

\begin{figure*}[t!]
\centerline{
\psfig{figure=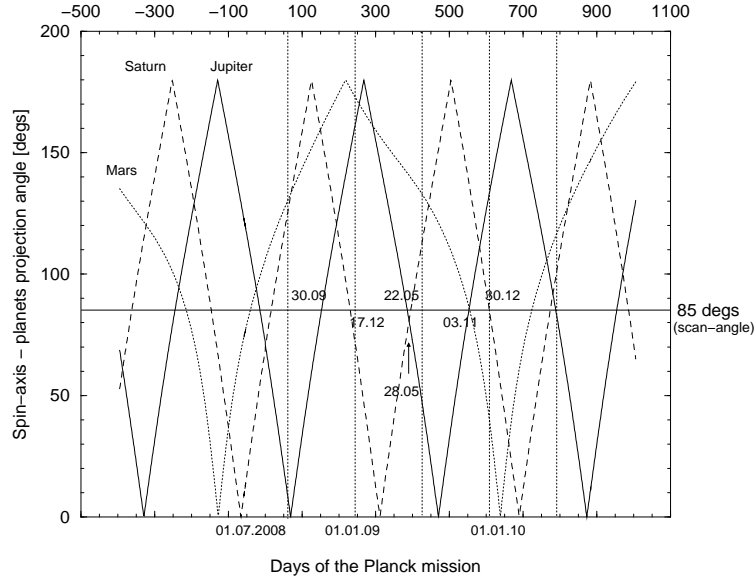,width=10cm,angle=-90}
}
\caption{Transit of Saturn, Jupiter and Mars through a scan angle
of $85\degr$. The dashed, solid and dotted curves represent the Saturn,
Jupiter and Mars paths, respectively.
The solid horizontal line demonstrates a scan angle of 85\degr.
Dates near the horizontal line correspond to
the Jupiter's and Saturn's crossings of the 85\degr scan angle.
} \label{aa1:Verkhodanov_n}
\end{figure*}

According to the present schedule, the
launch of the satellite is planned in June 30, 2008 and the flight to the
L2 point will take approximately six months. To construct the L2,
Earth, Jupiter and Saturn orbiting around the Sun, we use the planets
ephemerides  calculation procedure of the XEphem.3.7.2 software by
\cite{downey:Verkhodanov_n} (see Fig.~\ref{xephem:Verkhodanov_n}).

As mentioned above, the scan angle is assigned to be 85\degr.
Assuming
the beginning of the mission on the 31th of October 2008, in 15 months
Jupiter will cross the main beam direction 3 times on the dates of
22.05.2009, 03.11.2009 and 27.01.2010
(and on 09.12.2010)
with an accuracy of $\pm$1.5
day, mainly due to tilt and starting conditions, whereas Saturn will
cross the main beam 3 times on the dates of
17.12.2008, 31.05.2008 and 30.12.2009
(and in 2010, on 11.06.2010)
(Fig.~\ref{aa1:Verkhodanov_n}).

As shown in Fig.~\ref{aa1:Verkhodanov_n}, the antenna
beam shape calibrations using
Jupiter and Saturn can only be realized 6 ($3+3$) times during
the mission with a few different time
intervals between each planet crossing. These ``windows'' provide us with
the possibility of estimating degradation of the antenna main
beam shape from the long time frequencies (up to 15 months) to the short
frequencies (about 10 days).\footnote{The
interesting point of the track crossing between Jupiter and Saturn near
the scan angle on 26.05.2009 allows us to discuss the possibility
of  simultaneous observations of two planets in one scan.
However, to make such observations, we need to have a scan angle of
about 80\degr. Otherwise, these transits can be used to test the far
sidelobes of the beam.}
For continuous calibration of the main beam area, therefore, it is
necessary to use the method by \cite{beam:Verkhodanov_n} to reconstruct its
ellipticity
and orientation in the data analysis.
  One of the interesting dates in the planet observation is 03.11.2009,
when Jupiter and Mars could be observed during one day. Unfortunately, the
unpredictability of the Mars spectrum in the millimeter wavelength range
due to atmosphere storms
\cite{goldin:Verkhodanov_n}
does not allow us to  calibrate the far sidelobes. However, the mission
can provide interesting results for the study of Mars
\cite{planck:Verkhodanov_n}.

\subsection{Inclination of the Planet Orbits to the Ecliptic Plane}
We make our estimation of the planet transit for the ecliptic plane
projection. However, the planets have ecliptic latitude different from
zero.  The maximum inclinations are $\approx1\degr$ for Jupiter and
$\approx2\degr$ for Saturn (the detailed calculation for each planet crossing
is in Table~\ref{position:Verkhodanov_n}).
Here we note that this problem is not
essential for our case but is similar to the tilt projection problem
discussed below.

\begin{table*}[tbp]
\begin{center}
\caption{Position of the planets on the dates of crossing by Planck on
the ecliptic plane and
antenna temperatures
($T_a = \Omega_{\rm planet}/\Omega_{\rm beam} \cdot T_b$)
in transit via scan angle of 85\degr}
\label{position:Verkhodanov_n}
\medskip
\begin{tabular}{l|c|c|c|c|c}
\hline
 Planet&   Date	 & Size	  &   $T_a$ (mK)	    & Ecliptic &
Ecliptic  \\[-4pt]
       &	 &(arcsec)&~~~~30~~~~~100~~~545\,GHz& latitude &
longitude \\[-4pt]
       & 85\degr & &33\arcmin~~~~~~10\arcmin~~~~~~5\arcmin~&  &	 \\
\hline
Saturn	& 17.12.2008 & 17.9 & 7.534~~~86.98~~~291.2 & $+$1:58.0 &
171:27.2 \\
Jupiter & 22.05.2009 & 40.4 & 43.86~~~556.3~~~1672~ & $-$0:45.6 &
325:59.8 \\
Saturn	& 31.05.2009 & 17.9 & 7.534~~~86.98~~~291.2 & $+$2:07.8 &
164:57.5 \\

Jupiter& 03.11.2009 & 40.9 & 44.40~~~563.2~~~1693~ & $-$1:01.2 &
317:46.9 \\
Saturn & 30.12.2009 & 17.7 & 7.450~~~86.01~~~287.9 & $+$2:16.9 &
184:19.6 \\
Saturn & 17.06.2010 & 17.5 & 7.366~~~85.04~~~284.7 & $+$2:21.9 &
177:57.1 \\
Jupiter& 27.06.2010 & 40.4 & 43.86~~~556.3~~~1672~ & $-$1:16:03 &
2:10:53 \\
\hline
\end{tabular}
\end{center}
\end{table*}

According to the scan strategy proposals the spin axis can have a tilt
to the
ecliptic plane of about $10\degr$.
This means that the usual ecliptic projection of the scan angle will be
narrower and we shall observe Jupiter on other days with another
antenna temperature. The scan projection $\psi$ of the scan angle
$\alpha$ for the tilt angle $\phi$ can be calculated as
\begin{equation}
\tan \psi = \tan \phi \cos \alpha,
\end{equation}
which gives us $\psi=84.92^{\circ}$, for $\alpha = 85^\circ$ and
$\phi=10^\circ$. This difference is within $5\arcmin$, indicating
that the projections of the planet trajectories and the tilt of the axis
will not interfere with our estimates.

\subsection{Beam Descriptions and its Variations Due to Mirror
Degradation}


\begin{figure}[t!]
\centerline{
\psfig{figure=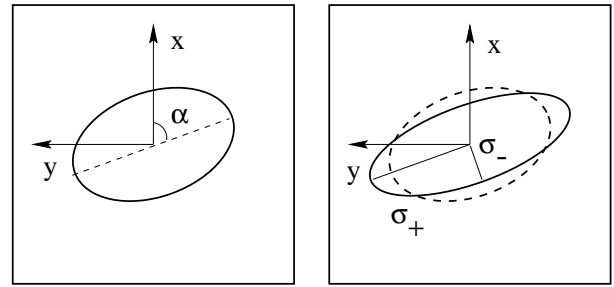,width=8cm,angle=-90}
}
\caption{Beam degradation parameters. The left panel illustrates the
orientation angle of the beam, and the right panel
shows $\sigma_{-}$ and $\sigma_{+}$, the semi-minor and semi-major axes of the
elliptical main beam.}
\label{degradation:Verkhodanov_n}
\end{figure}

The Planck ``antenna beam'' is usually referred to as a physical
model of the antenna response and its ground-based verification before
the flight, and the {\it in-flight} antenna beam as the beam
reconstructed
during the flight, which is crucial
for possible beam degradation estimation. The in-flight antenna beam
plays
a significant role in the $C_\ell$ estimation as well as in the
extra-galactic point source extraction \cite{tophat:Verkhodanov_n}.

Physics optics calculations have shown that the main beam is roughly
elliptical \cite{elliptical:Verkhodanov_n,burigana2001:Verkhodanov_n},
so we can approximate the
antenna pattern as a bivariate Gaussian beam. The geometrical property
of the beam in the time domain can be described as follows.
We denote by
$x_0$ and $y_0$ the position of the beam in a coordinate system fixed
to the detector with $x$ in scan direction and $y$ perpendicular to
$x$ and the beam axis.
Then the beam shape can be written as
\begin{equation}
B_t(\vec{x}-\vec{x}_t)=
\exp\left[-\frac{1}{2}({\sf RU})^{\rm T} {\sf D}^{-1}({\sf RU})\right],
\end{equation}

\noindent with

\begin{equation}
{\sf U}=\left(
\begin{array}{c}
    x-x_0 \\
    y-y_0
\end{array}
\right),
\end{equation}
where ${\sf R}$ is the rotation matrix which describes the inclination of
the elliptical beam,
\begin{equation}
{\sf R}=\left(
\begin{array}{rr}
       \cos \alpha  &  \sin \alpha  \\
      -\sin \alpha  &  \cos \alpha
\end{array}
\right),
\end{equation}
with $\alpha$ being the orientation angle between the $x$ axis and the
major axis of the ellipse. The ${\sf D}$ matrix denotes the beam
width along the ellipse major axis, which can be expressed as
\begin{equation}
{\sf D}=\left(
\begin{array}{cc}
      \sigma^2_{+} &	 0	    \\
      0	       &  \sigma^2_{-}
\end{array}
\right).
\end{equation}

\begin{figure}[t!]
\centerline{
\psfig{figure=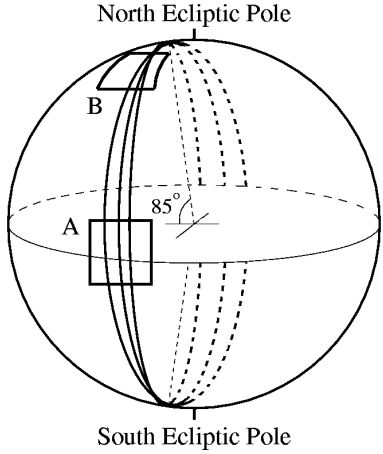,width=5cm}
}
\caption{The Planck scan strategy. The square $A$ is located at the
parallel scan area, where the planets can be found. The square $B$ has
crossing(s) of the circular scans.}
\label{scan:Verkhodanov_n}
\end{figure}

If, during the routine operation, the mirror surfaces are slightly
perturbed (deformed), it is necessary to conduct a detailed
investigation of the corresponding degradation of the
antenna beam shapes, using Jupiter's and Saturn's transits as suggested by
\cite{burigana2001:Verkhodanov_n}\footnote{As we will see
in the next section, the Jupiter's flux
fluctuation could mimic  the in-flight beam degradation effect as well.}.
In
general, beam degradation can be described by 3 parameters as
illustrated in Fig.~\ref{degradation:Verkhodanov_n}.
First of all, the mean beam
width is a function of time,
\begin{equation}
\frac{1}{2}(\sigma_{-}^2+\sigma_{+}^2)\equiv \sigma^2 = \sigma^2(t),
\end{equation}
where $\sigma_{-}$ and $\sigma_{+}$ are the minor and major axes of
the elliptical main beam. The orientation angle $\alpha$
between the scan direction and the major axis of the elliptical main
beam can also be a function of time, i.e.,
\begin{equation}
\alpha\equiv \alpha(t).
\label{eq:iangle:Verkhodanov_n}
\end{equation}
and so can the ellipticity ratio of the beam
\begin{equation}
\frac{\sigma_{+}}{\sigma_{-}} = \rho \equiv \rho(t).
\label{eq:eratio:Verkhodanov_n}
\end{equation}

We can use these 3 parameters as the indicators of the degradation
level of the in-flight antenna beam.

\section{PLANET TRANSITS AND THE PIXEL DOMAIN}

Following \cite{burigana2001:Verkhodanov_n},
we can specify the in-flight Planck
antenna beam shape model by using Jupiter as a ``standard candle'' for
calibration. For the Planck (both LFI and HFI) frequency ranges we
can model the Jupiter's contribution
to the resulting $\Delta T(\mbox{\boldmath $r$})$
 sky temperature in some direction $\mbox{\boldmath $r$}$ as
\begin{equation}
{\textstyle
\Delta T_{\rm J}(\mbox{\boldmath $r$})=\frac{S_{\rm J}(\nu,
t)}{2k}\left(\frac{hc}{kT_{\rm CMB}}\right)^2
\left[\frac{2 \sinh(\eta/2)}{\eta^2}\right]^2\delta(\mbox{\boldmath $r$}_{\rm J},\mbox{\boldmath $r$}),
\label{eq:eq1:Verkhodanov_n}}
\end{equation}
where \mbox{\boldmath $r$} and $\mbox{\boldmath $r$}_{\rm J}$ are
the unit vectors in the
corresponding direction on the sky and Jupiter's location in a given
coordinate system, respectively, $\delta(\mbox{\boldmath $r$}_{\rm J},\mbox{\boldmath $r$})$
 is the
Dirac delta function, $S_{\rm J}$ is the Jupiter's flux, $T_{\rm
CMB}$, $h$, $k$ and $c$ are the CMB temperature, Planck constant,
Boltzmann constant, and speed of light, respectively, and
$\eta=h\nu/kT_{\rm CMB}$.

Each observed time-ordered subscan $m^{i}_t$ that includes Jupiter's
image is related to $\Delta T_{\rm J}(\mbox{\boldmath $r$})$ from
Eq.~(\ref{eq:eq1:Verkhodanov_n}) through a convolution with the antenna beam
function $B(\mbox{\boldmath $r$},\mbox{\boldmath $r$}^{'})$
(see \cite{beam:Verkhodanov_n})
\begin{equation}
y^{i}_t= d^{i}_t +n^{i}_t\,,
\label{eq:eq2:Verkhodanov_n}
\end{equation}
where
\begin{equation}
d^{i}_t=     \Delta T_{\rm J}[\mbox{\boldmath $r$}(t)]\otimes
B[\mbox{\boldmath $r$}(t^{i}),\vec{r}(t)],
\label{eq:eq3:Verkhodanov_n}
\end{equation}
where $\otimes$ denotes convolution,
and $n^{i}_t$ now is the CMB signal plus all
the foregrounds and the instrumental noise.
 The index $i$ marks the $i$-th subscan with
the same orientation of the spin axis of the satellite.

Using all $i \in[1,N]$ subscans, where $N$ is the total number of
the subscans,
we can define the circular scan
as some linear transformation of the $d^{i}_t$:
\begin{equation}
d_t={\sf A}\, {\sf d}^i,
\label{eq:eq31:Verkhodanov_n}
\end{equation}
where ${\sf d}^i=\left\{d^{i}_t\right\}$ is the data vector, and ${\sf
A}$ is the matrix of the transformation.
Eq.~(\ref{eq:eq31:Verkhodanov_n}) gives
us the relation of a single circular scan for a fixed orientation of the
spin axis. For
simple summation of the subscans we obtain $d_t=(1/N)\sum_i d^{i}_t$.
Below we
shall
use the circular scan as a basic element for
the map-making algorithm taking into account that the variance of the
instrumental noise for such a scan is expected to be $\sim N^{-1/2}$
times smaller than for each subscan, if the instrumental noise
is pure white noise. For all circular scans we can define the vector
of the time-ordered data ${\sf y}={\sf M}\, {\sf s} + {\sf n}$, where
${\sf M} $ is the corresponding map-making matrix, ${\sf s}$ denotes
the pixelized map and ${\sf n}$ is the noise vector
\cite{tegmark1997:Verkhodanov_n}.
It is worth noting that for the in-flight antenna
beam shape reconstruction by using Jupiter's and Saturn's images, we do
not need to construct whole-sky
maps,
because the $-40$\,dB limit of the
expected Planck antenna beam shape corresponds to an angular scale
$\theta_{\rm fs}\sim 5$ degrees at 30\,GHz LFI, \footnote{For LFI+HFI
frequency range this scale corresponds to the minimum.}
and for that purpose we can use the
flat-sky approximation centered around t
he Jupiter's image (Fig.~\ref{scan:Verkhodanov_n}),
and apply the method by \cite{beam:Verkhodanov_n}.
Furthermore, this assumption allows us to use a
linear map-making algorithm (see \cite{tegmark1997:Verkhodanov_n}), which is
similar, for example, to the COBE pixelization scheme.\footnote{Note
that for estimations of the in-flight beam distortions caused by $1/f$
noise, foreground contaminations and so on,
\cite{burigana2001:Verkhodanov_n} have
to use the whole-sky map for the $1/f$ noise removal.}

The signal in each pixel of the map ${\sf s}$ is then
\cite{tegmark1997:Verkhodanov_n}
\begin{equation}
{\sf s}={\sf W}{\sf y},
\label{eq:eq4:Verkhodanov_n}
\end{equation}
where ${\sf W}$ is the corresponding matrix, which depends on the scan
strategy of the Planck experiment. For example,
for the simplest COBE pixelization we can use
${\sf W}=\left[{\sf M}^{\rm  T}{\sf N}^{-1}{\sf M}\right]^{-1} {\sf
M}^{\rm T}{\sf N}^{-1}$, where ${\sf N}=\langle \mbox{\boldmath $nn$}^{\rm
T}\rangle$ is the noise covariance matrix. \footnote{Note that we can
use any modification of pixelization without loss of information, e.g.
GLESP \cite{glesp1:Verkhodanov_n,glesp2:Verkhodanov_n} or
HEALPix \cite{healpix:Verkhodanov_n}.}

Let us go back to the single circular scan. As seen from
Eq.~(\ref{eq:eq1:Verkhodanov_n}) and Eq.~(\ref{eq:eq3:Verkhodanov_n}),
for the simple average of
the subscans (${\sf A}\rightarrow1/N\sum_i$) the Jupiter's image after
beam convolution in a circular scan should be
\begin{equation}
d_t= \frac{1}{2kN}
\left[\frac{2 \sinh(\eta/2)hc}{kT_{\rm CMB}\eta^2}\right]^2
\sum_i\sum_{t^{'}_i}
S_{\rm J}(\nu, t^{'}_i)B_t({\bf \theta}_i),
\label{eq:eq41:Verkhodanov_n}
\end{equation}
where ${\mbox{\boldmath $\theta$}}_i=\mbox{\boldmath $r$}^{'}(t^{'}_i)-\mbox{\boldmath $r$}(t_{{\rm J},i})$,
$t_{{\rm J},i}$ is the Jupiter's location in the $i$-th subscan and
$B_t$ denotes the beam shape in each subscan. As one can see from
Eq.~(\ref{eq:eq4:Verkhodanov_n}) and Eq.~(\ref{eq:eq41:Verkhodanov_n}),
in the pixelized map the
 pixels containing Jupiter's image are related to the $B_t$ and $S_{\rm
J}(\nu, t^{'}_i)$ and can be denoted as
follows,
\begin{equation}
{\sf s} \propto {\sf W}S_{\rm J}(\nu, t^{'}_i)B_t(\mbox{\boldmath $\theta$}_i).
\label{eq:eq42:Verkhodanov_n}
\end{equation}

\begin{figure*}
\centerline{
\psfig{figure=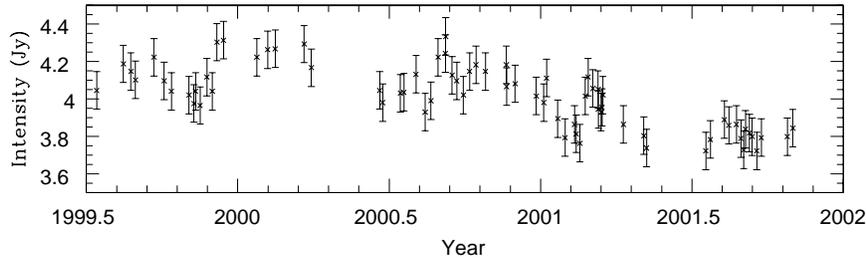,width=12cm}
}
\caption
{Variability of the flux density of synchrotron radiation
at 4.04 AU from Jupiter at
13 cm wavelength (2.3\,GHz) (re-produced from \cite{bolton:Verkhodanov_n}).}
 \label{variability:Verkhodanov_n}
\end{figure*}

Therefore, for the definition of the antenna beam shape in the pixel
domain, we can specify some possible sources of uncertainties from
Eq.~(\ref{eq:eq42:Verkhodanov_n}).

Before focusing on the two categories of variabilities in beam
calibration in the next two sections, we would like to briefly mention
the uncertainty which is related to the location of Jupiter,
\begin{equation}
\mbox{\boldmath $r$}(t_{{\rm J},i})=\overline{\mbox{\boldmath $r$}}(t_{{\rm J},i}) +
\Delta {\mbox{\boldmath $r$}}(t_{{\rm J},i})=
\overline{\mbox{\boldmath $r$}}(t_{{\rm J},i})(1+\delta_r), \label{eq:eq5:Verkhodanov_n}
\end{equation}
where $\overline{\mbox{\boldmath $r$}}(t_{{\rm J},i})$ indicates
the average location
and
$\Delta \mbox{\boldmath $r$}(t_{{\rm J},i})$ corresponds to the
fluctuation of the
Jupiter's
location. Generally speaking, this source is related to the pointing
accuracy
of the Planck experiment. It is natural to assume that
$\langle \Delta \mbox{\boldmath $r$}(t_{{\rm J},i})\rangle = 0$, but $\langle
|\Delta \mbox{\boldmath $r$}(t_{{\rm J},i})|^2\rangle \neq 0$.

\begin{figure}
\centerline{
\psfig{figure=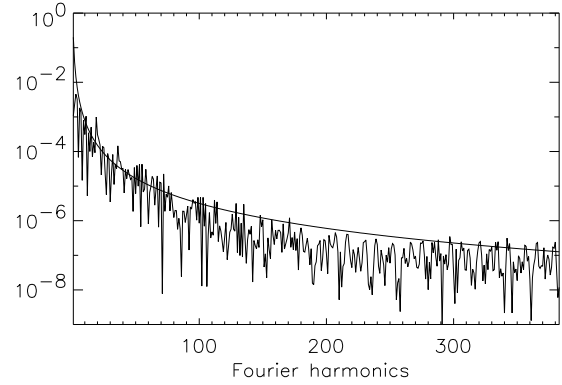,width=8.5cm}
}
\caption
{The power spectrum of the flux density
from Jupiter at the 13\,cm wavelength, which is produced by linear
interpolation in the parts of the intervals in
Fig.~\ref{variability:Verkhodanov_n}
where data are not available. The fitted curve is described by a power
law $P(k)\propto k^{-2.4}$.}
\label{powerspectrum:Verkhodanov_n}
\end{figure}

\section{VARIABILITY OF THE JUPITER's FLUX ON THE BEAM CALIBRATION}

The fluctuations of the Jupiter's flux can be crucial for the in-flight
antenna beam shape reconstruction scheme. The temporal variations in the
Jupiter's flux can be expressed as the constant flux $\overline S_{\rm
J}(\nu)$, and a fluctuating part $\Delta S_{\rm J}(\nu, t)$,
\begin{equation}
S_{\rm J}(\nu, t)=\overline S_{\rm J}(\nu) + \Delta S_{\rm J}(\nu, t)=
\overline S_{\rm J}(\nu)[1+\delta_S(t)].
\label{eq:eq43:Verkhodanov_n}
\end{equation}

Returning to Eq.~(\ref{eq:eq4:Verkhodanov_n}) and
Eq.~(\ref{eq:eq41:Verkhodanov_n}) in order
to define the beam shape properties in the pixel domain, we will
 assume $\langle \Delta \mbox{\boldmath $r$}(t_{{\rm J},i})\rangle = 0$
(so that in
Eq.~(\ref{eq:eq5:Verkhodanov_n}) $\delta_r=0$).
The pixelized beam can be obtained
from the subscans including Jupiter's image as
follows \cite{wu:Verkhodanov_n}:
\begin{equation}
B_p({\mbox{\boldmath $\gamma$}})= \sum_{t\in p}\sum_i \sum_{t^{'}_i} {\sf W}
[1+\delta_S(\nu, t^{'}_i)]B_t({\mbox{\boldmath $\theta$}}_i),
\label{eq:eq44:Verkhodanov_n}
\end{equation}
where ${\bf \gamma}$ is the angle between
the position of pixel which corresponds to
the Jupiter's location in a map and position of each different
pixel. Possible variation of the Jupiter's
flux produces an additional source
of
peculiarities in the pixelized beam shape definition proportional to
$1+\delta_S(t)$.

\subsection{Characteristic Time Scales}

There are 3 characteristic time scales related to the Planck
scan strategy. For each subscan of 1 r.p.m. the time scale is
 $ T_{\rm sub}\simeq 1$ minute. The next time scale
is that for a circular scan $T_{\rm cir}=60$ minutes, which is the
time interval for data accumulation in one circular scan with a fixed
orientation of the spin axis. In terms of order of magnitude, $T_{\rm
cir}$ scale can be used for estimation of the characteristic time
scale for the signal variation in one pixel around the main beam
area. Another time scale is related to the scale of the far sidelobes
$T_{\rm FS} \simeq \theta_{\rm FS}/1^{\circ}$ days, where $\theta_{\rm
FS}$ is the angular measure subtended by the far sidelobes of the
beam. For example, for the LFI 30\,GHz channel, the threshold of $-30$
dB subtends the angular scale $\theta_{\rm FS} \simeq 1.5$ degrees
\cite{burigana2001:Verkhodanov_n} and thus $T_{\rm FS} \simeq 1.5$ days. The
high-frequency fluctuations of the Jupiter's flux which corresponds to
the time scales $T_{\rm sub}$, $T_{\rm cir}$ and $T_{\rm FS}$ are thus
very important for the in-flight antenna beam shape reconstruction and
may require more detailed investigations, for example by ground-based
telescopes. These time scales also indicate that all irregularities of the
Jupiter's flux for $ T \ge T_{\rm FS}$ correspond to long-term
variations and could mimic a beam shape degradation effect.

Unfortunately, we do not know exactly the properties of temporal
variations in the Jupiter's flux in the Planck frequency range 30--857
GHz. According to the information  available in the
literature, the Jupiter's flux variability at a frequency of 2.3\,~GHz,
nearest to
LFI frequency range, is
 related to synchrotron emission from the Jupiter's
magnetic belts.
Fig.~\ref{variability:Verkhodanov_n} shows the variation of
the flux density of the
synchrotron radiation from Jupiter at 2.3\,GHz, reproduced
from \cite{bolton:Verkhodanov_n}. The interval of measurement is one day, so
any fluctuation shorter than 1 day is yet to be measured. In
Fig.~\ref{powerspectrum:Verkhodanov_n} we show the power spectrum of
Fig.~\ref{variability:Verkhodanov_n}, which is produced from
linear interpolation in
the part of the intervals where data are not available.

The temporal variation that can significantly distort the beam
reconstruction
for the 30\,GHz channel is between 2.4 hours and 1.5 days, the Planck
crossing time of Jupiter. The variation period of the Jupiter's flux shorter
than 2.4 hours may be smeared out after pixelization in this channel
because of the scan strategy and beam shape properties. For a variation period
longer than 1.5 days, the distortion is much less.

\begin{figure}[t!]
\centerline{
\psfig{figure=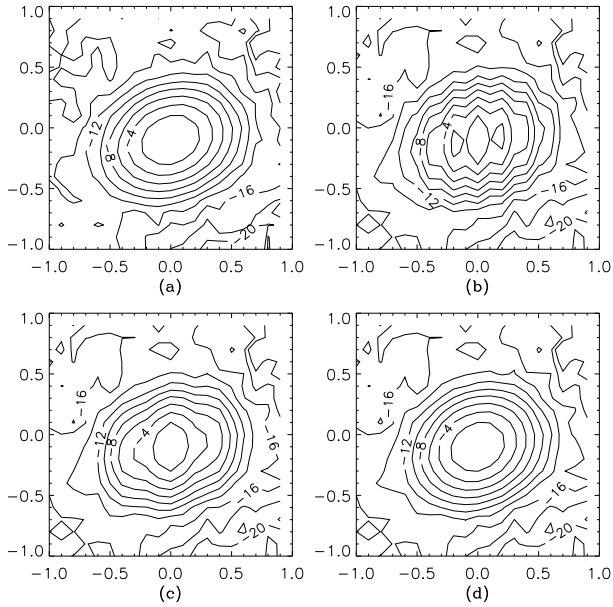,width=8.5cm}
}
\caption{Simulation of beam reconstruction with variation of Jupiter's
flux. We simulate  30\,GHz LFI channel (FWHM$=33\arcmin$ with an
ellipticity ratio =1.3) with $\sigma_{\rm CMB}$ and pixel noise
$\sigma_{\rm pix}$ equal to $3 \times 10^{-5}$ and $8 \times 10^{-6}$,
respectively. Panel (a) is without Jupiter's flux variation, (b) with
Jupiter's flux varying with a period of 4.8 hours, (c) 10 hours (the
rotation period of Jupiter) and (d) 1.5 days (the Planck crossing
time of Jupiter). The contour lines are annotated in dB.}
\label{flux:Verkhodanov_n}
\end{figure}

In Fig.~\ref{flux:Verkhodanov_n} we show simulated beam reconstructions in the
30\,GHz LFI
channel with possible Jupiter's flux variations. According to
Eq.~(\ref{eq:eq43:Verkhodanov_n}), if $\delta_S(t)$ is a random process,
in the
Fourier domain it can be characterized by a power spectrum.
We will assume that for Jupiter atmospheric emission the power
 fluctuations $\delta_S(t)$ should have a form
$P(\omega)\simeq a\omega^{-n}+b$, where $a$ and $b$ are constant.
For illustration we will use the same values of $a$ and $b$ parameters
as for the synchrotron emission discussed above
 (Fig.~\ref{powerspectrum:Verkhodanov_n}).
The flatness of the power spectrum at large Fourier modes in
Fig.~\ref{powerspectrum:Verkhodanov_n} allows us to assume
the same amplitude of the
variation for different periods. Panel (a) is the reconstructed image
without variation in the Jupiter's flux, (b), (c) and (d) are with flux
fluctuation of the variation period equal 4.8 hours, 10 hours, and 1.5
days, respectively. The period of 10 hours corresponds to that of Jupiter's
rotation. The amplitude of the variations is assumed to be 20\%. We can
clearly see that the orientation of the main beam changes because of the
fluctuation of the Jupiter's flux.

\subsection{Millimeter Spectra of the Planets and Their Variations at
Different Frequencies}

\begin{figure}[t!]
\centerline{
\vbox{
\psfig{figure=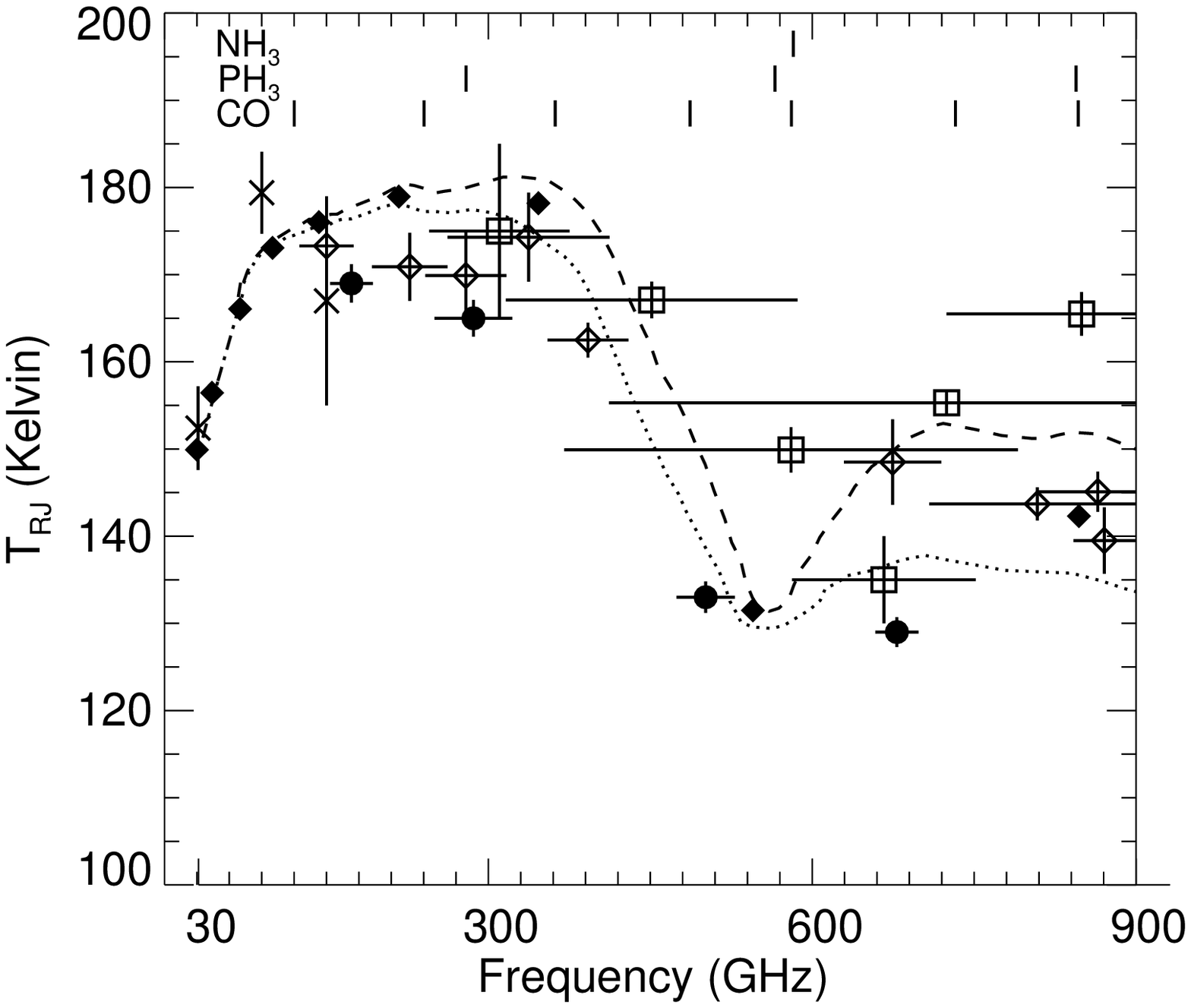,width=7.5cm}
\psfig{figure=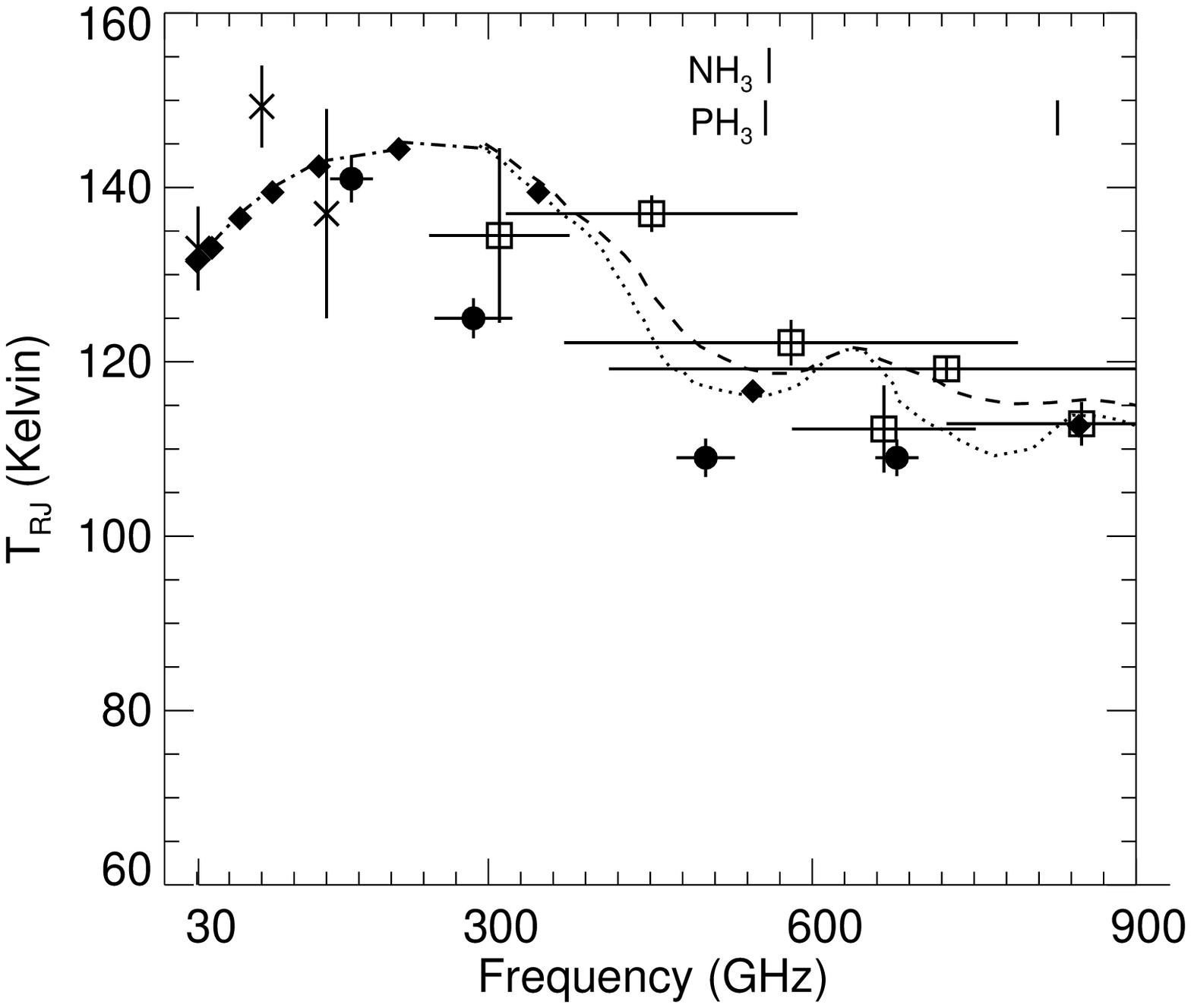,width=7.5cm}
}
}
\caption{The millimeter and sub-millimeter spectrum of Jupiter
(top) and Saturn (bottom) (reproduced from \cite{goldin:Verkhodanov_n}).
The filled diamonds denote the fluxes calculated in this work according
to the model at the 9
observing frequencies in the Planck experiment. Note that the
first strong dips in the spectra coincide around Planck 545 GHz
observing channel.
The figure of spectrum was prepared using
\cite{goldin:Verkhodanov_n} (filled circles), \cite{griffin:Verkhodanov_n}
(diamonds),
\cite{hildeb:Verkhodanov_n} (squares),\cite{ulich:Verkhodanov_n} (crosses).}
\label{jupsat:Verkhodanov_n}
\end{figure}

To estimate the effects of the flux density variation of the planets,
we have to look at their radio spectra. The total Jupiter's radio
spectrum
consists of the following two components \cite{burke:Verkhodanov_n}: a low
radio-frequency part, which is related with the synchrotron emission from
energetic
electrons spiraling in the Jupiter's magnetic field, and a high
radio-frequency part, which corresponds to the thermal atmospheric emission.

The synchrotron emission dominates in the frequency range $\nu\le 10$
GHz while at  $\nu\ge 30$ GHz the Jupiter's radio flux is determined by
the atmospheric emission.
First of all, let us  describe the contribution of the synchrotron emission
and its variability to the $ \delta_S(t_{\rm J})$ parameter for the
 30\,GHz LFI channel.
Recently simultaneous observations with the Cassini spacecraft, the
Gallileo
spacecraft and the VLA in the centimeter wavelength range have been made
\cite{gurnett:Verkhodanov_n,bolton:Verkhodanov_n}.
As is shown by \cite{gurnett:Verkhodanov_n}, the
  Jupiter's magnetosphere is strongly affected by the solar wind. When
interplanetary
shocks propagate  from the Sun and reach Jupiter, they
compress and re-configure the magnetosphere, producing a strong
magnetic field and electron acceleration.
The $\sim$10\% variation with a variability of about 0.5\,Jy per month
of the 13\,cm flux from Jupiter observed by
\cite{bolton:Verkhodanov_n}, and shown
in Fig.~\ref{variability:Verkhodanov_n}, may
most likely be due to such an effect. This value can be used as
an upper limit in our estimations.
For the Planck antenna beam shape reconstruction we must,
as shown earlier,  know the
variability of the Jupiter's flux on a time scale of $T_{\rm cir}\sim
1\div2$ days, to obtain the lower limit of the variation for the
Planck frequency range 30--857~GHz. Obviously, the synchrotron emission
is
important, in principle, only for 30\,GHz channel and it determines the
lower limit of the Jupiter's flux variation, if the atmosphere emission
does not produce any fluctuations of the flux in the 30\,GHz band.
We plot in Fig.~\ref{powerspectrum:Verkhodanov_n}
the power spectrum of the 13\,cm synchrotron flux variability.
From this spectrum we find the limit $\Delta_{\rm synch}\sim $10\% per
day.

If we argued that the same amplitude of the synchrotron emission also
occurred at the 1\,cm wavelength and used this value in our simulations,
it would give the lower limit of the
flux variation due to the synchrotron emission at 30\,GHz of $\delta_{\rm
synch}
\sim (T^{\rm synch}_b/T_b)\Delta_{\rm synch} \simeq 10^{-3}$,
where $T_b$ is the brightness temperature corresponding to the total
planet flux and $T^{\rm synch}_b$
that corresponding to the
synchrotron emission.
Thus, we can conclude that the variation of the synchrotron emission
at $\nu=30$~GHz is not important for the antenna beam shape
reconstruction for the whole range of interest ($\ge -60$~dB). However, it
is necessary to obtain additional observational data on
the intrinsic atmospheric emission. \footnote{We would like to argue
that it is natural to expect a non-zero fluctuation from the atmospheric
emission. Partly our assumption is based on the millimeter and
sub-millimeter spectrum of the atmospheric emission measurements at
30--857~GHz.}

Detailed studies  by \cite{goldin:Verkhodanov_n} of the millimeter and
sub-millimeter spectra of Jupiter and Saturn have shown
(see Fig.~\ref{jupsat:Verkhodanov_n})
that there are peculiarities in the spectra in this wavelength range.
The two model spectra shown in Fig.~\ref{jupsat:Verkhodanov_n} are
from \cite{griffin:Verkhodanov_n}, 
using different physical parameters such
as the size of ${\rm NH}_3$ clumps and the particle and gas scale
heights ratio.
Temperatures of Jupiter and Saturn estimated at the
corresponding Planck observing frequencies  are shown by
filled diamonds	 in Fig.~\ref{jupsat:Verkhodanov_n}. The first strong dip
on both spectra almost coincides with the observing frequency 545\,GHz
near the
570\,GHz
${\rm NH}_3$ and ${\rm PH}_3$ resonances.
Estimated brightness temperatures with an accuracy of about
10\%  are given in Table~\ref{temperature:Verkhodanov_n} (see also
Table~\ref{position:Verkhodanov_n}).

\begin{table}[tbp]
\caption{Brightness temperature $T_b$ of Jupiter and Saturn  at
Planck observing frequencies with an approximately 10\% accuracy}
\label{temperature:Verkhodanov_n}
\begin{tabular}{c|r|c|c|c}
\hline
 $\nu$	& $\lambda$ &  beam   &	 $T_{\rm Jupiter}$ & $T_{\rm Saturn}$ \\ [-4pt]
  (GHz)	  &    (mm)	& (arcmin)  &	 (K)	   &	(K)	 \\
\hline
   30	&    10.00   &	 33.0	 &    152     &	   133	  \\
   44	&    6.82    &	 24.0	 &    158     &	   135	  \\
   70	&    4.29    &	 14.0	 &    167     &	   138	  \\
  100	&    3.00    &	 10/9.2	 &    173     &	   141	  \\
  143	&    2.10    &	 7.1	 &    176     &	   144	  \\
  217	&    1.38    &	 5.0	 &    179     &	   146	  \\
  353	&    0.85    &	 5.0	 &    178     &	   141	  \\
  545	&    0.55    &	 5.0	 &    133     &	   118	  \\
  857	&    0.35    &	 5.0	 &    145     &	   114	  \\
\hline
\end{tabular}
\end{table}
Unfortunately, we do not
have information about variability of the Jupiter's and Saturn's fluxes
in the range 30 to 857\,GHz, which determines the accuracy of the beam
shape reconstruction.
Some naive expectation of a possible variability
in the frequency range of interest for the Planck mission
could be related with the observed  20\%
deviations of the Jupiter's and Saturn's temperature from the pure
black body law $T(\nu)$=const. For Jupiter, this  20\% deviation
allows us to expect that some process leading to such kind of
variations can be variable in time at the same level and have the
characteristic time scale close to the period of Jupiter's rotation
(i.e. $\simeq 10$ hours).  This problem needs an additional and more
detailed
investigation by using large ground-based radio telescopes in order
to measure possible variation of the Jupiter's and Saturn's fluxes
in the Planck frequency range.
Recent explorations of Jupiter with the radio telescope RATAN-600
have shown  stability of the Jupiter's flux at the level of
0.1\%  at the frequency of 30\,GHz for
34 days of observations \cite{par:Verkhodanov_n}.
Such type observations will be necessary during ``in-flight'' beam
calibration of the
Planck mission.

\subsection{Expected Polarization of the Flux}

One of the main goals of the Planck mission is the CMB
polarization measurements. First of all, we would like to point out that a low
limit
of  polarization of the Jupiter's flux at 30\,GHz exists, which is related
to synchrotron emission.

According to \cite{cortiglioni:Verkhodanov_n}, the polarization level $\Pi$
of synchrotron radiation is related to the spectral index $\beta$
($T_{\rm synch} \propto \nu^{\beta}$) as
\begin{equation}
\Pi = \frac{3\beta + 3}{3\beta + 1},
\end{equation}
which gives $\sim 10 \div$75\% for different values of $\Pi$.
The total polarized flux of Jupiter and Saturn is \mbox{$\sim T_{\rm
planet}\Pi$.}
This fact creates a pre-condition to use these planets for polarized
antenna beam shape calibration.

Using the value of $\beta=-1.26$ for $\nu >13.6$\,GHz for the spectral index
of the Jupiter's synchrotron flux \cite{bolton:Verkhodanov_n}, one can obtain
$\Pi_{\rm Jupiter}$=28\%, indicating that the polarized part of the
total flux can reach around 0.3\% in the 1\,cm wavelength range
(see Table 2).

\section{EFFECTS FROM THE STRATEGY OF THE OBSERVATIONS}

\subsection{Variations of Planet Antenna Temperatures Versus
Distance to the Planets}
From the dates of crossing the scan angle by the planets,
one can calculate the corresponding
distance and hence the angular sizes of the planets.
When the object size is sufficiently smaller than the solid angle of the
beam,
the antenna temperature $T_a$ of the planet is given by
\begin{equation}
    T_a = T_b \frac{\Omega_{\rm planet}}{\Omega_{\rm beam}},
\label{eq:ta:Verkhodanov_n}
\end{equation}
where $T_b$ is the brightness temperature of the planet
(Table~\ref{temperature:Verkhodanov_n}), $\Omega_{\rm planet}$
is the solid angle of
the planet in observation (angular size in steradian), and
$\Omega_{\rm beam}$ is the solid angle of the beam calculated with a
simple
approximation by a Gaussian shape
\begin{equation}
B=\exp\left(-\frac{\theta^2}{2\sigma^2}\right), \label{eq:beam:Verkhodanov_n}
\end{equation}
where  $\sigma \equiv \sqrt{(\sigma_{-}^2+\sigma_{+}^2)/2}=\theta_b /
2.355$ and $\theta_b$ is the FWHM of the main beam.
The results of calculation of the antenna temperatures for
the three frequencies 30, 100 and 545\,GHz with the corresponding
FWHM sizes of 33, 10 and 5 arcmin are given in Table~\ref{position:Verkhodanov_n}.

\subsection{Scan Strategy and Peculiarities of the In-flight HFI Beam
Reconstruction}

In this subsection we would like to focus on the discussed
scan strategy of the Planck mission and its influence on the
in-flight antenna beam shape reconstruction using Jupiter's and Saturn's
transits. According to the Planck mission requirement, the FWHM for
the 10 LFI + HFI channels is shown in Table~\ref{temperature:Verkhodanov_n}.
Let us
concentrate on the beam shape properties above $- 30$\,dB
for all LFI + HFI  channels.

According to the scan strategy (see Fig.4), the orientation of
the telescope spin axis during one hour (i.e. 1 r.p.m. of spin for 60
sub-scans) of observations should be stable: the orientation of the
60th sub-scan is parallel to that of the 1st sub-scan at the moment
$t_0$ when a given circular scan starts to be measured. At the end of the
60th sub-scan, the spin axis (and the optical axis) should change
its orientation by $2\farcm5$ in the ecliptic plane (re-pointing).
Thus, during Jupiter's and Saturn's transits the highest resolution scale
from which the images of Jupiter and Saturn  can be recovered in
the pixelized map is $2\farcm5$ on one side
(as the sizes of both the planets are
less than 1\arcmin).

Due to the pixelization scheme, however,
we will face the following two situations
concerning  the higher frequency channels:  good and  bad cases of the
planet transit. The good case is where the planets are caught by the
beam peak just after re-pointing and have a maximum signal on a
circular scan. The probability of such a case is low. The bad case is
where the planets bypass the beam maximum so that the point of maximum
of the planet flux is missed by both of the neighboring circular scans.
In
terms of the map-making algorithm it means that the point of the
maximum of the Jupiter's flux is formally shifted away from the center of the
corresponding pixel and the signal in the surrounding pixels must be
asymmetrical. This asymmetry can be removed using the expected $10\arcsec$
pointing accuracy if there are no temporal variations in the Jupiter's flux
or no mirror degradation effect.

Thus, the map-making algorithm should reflect directly the scan
strategy and the position of Jupiter and Saturn
(see Fig.\ref{xephem:Verkhodanov_n} and \ref{scan:Verkhodanov_n}). Using
the $-30$\,dB threshold, we can
estimate the number of pixels, which manifests the beam shape in the map
for
each frequency channel. For the simple Gaussian approximation (see
Eq.(~\ref{eq:beam:Verkhodanov_n})) we get
\begin {equation}
N_{\rm pix}(\nu)\simeq 9 \left(\frac{{\rm FWHM}(\nu)}{2\farcm5}\right)^2.
\label{eq:b:Verkhodanov_n}
\end{equation}
As one can see from Eq.~(\ref{eq:b:Verkhodanov_n}),
$N \approx 1570$ pixels in the 30\,GHz
 channel (FWHM $\simeq 33\arcmin$), while $N \approx 36$ pixels only for
the 217~GHz (and higher) channels (FWHM $\simeq 5\arcmin$), which is
obviously
not enough to accurately determine these beam shape ellipticity,
particularly if the ellipticity is no larger than 1.2.

\section{CONCLUSIONS}

In summary, regarding the issues related to using transits of planets, such
as Jupiter and Saturn, as a  method of calibration of the in-flight beam
shape, we
conclude the following:
\begin{itemize}
\item
    The high accuracy of the $C_\ell$ estimation by Planck will
    require the main beam estimation with an error of 1\%, which
    implies that we need to measure the Jupiter's flux variation $\delta
    B/B=\delta_{\rm S}$ to the same level.
\item
    For observations in the LFI frequency range, e.g. $30\div100$\,GHz,
     we	 have a limit of possible variations of the Jupiter's flux
    $\le 0.1$ for the $-30$~dB threshold of the beam. In addition, we
need
    to measure the variations of the Jupiter's flux by using
    ground-based radio telescopes.
\item
    When~~ observing~~ at~~ high~~ frequencies (217$\div$857\,GHz) we may have
    problems during observation of the planets such as missing a
    target due to the narrow beams.
\item
    Practically, during the mission ($\approx 15$ months), we will be
    able to test the main beam about 3 times down to $-23.5$\,dB
    by using the Jupiter's flux and 2 times by Saturn.
    This implies
    that in the case of calibrating the beam degradation effect in
    intervals shorter than 3 months we have to use the method by
    \cite{beam:Verkhodanov_n}.
\item
    The close (in time interval) transit of the planets will
    enable us to check a high--frequency component of the beam
    degradation.
\item
    Neither Jupiter nor Saturn makes it possible  to
    test the far sidelobes.
\item
    Galactic synchrotron and dust emissions appear at a level of 0.18\,mK.
\item
    The possible degradation effect could be important at the same, as
    the Jupiter's flux, level of variation, $\sim 1\div 10$\%, for the
    beam width.
\item
    At low frequencies (30\,GHz) of the mission,
    the stability of the Jupiter's flux can be monitored with the RATAN--600
    radio telescope. Such a possibility was demonstrated in investigation
    of the planet radio flux whose stability was checked during
    a month at a level of 0.1\% \cite{par:Verkhodanov_n}.
\end{itemize}

We also note that  calibrations by Jupiter and Saturn,
together with the method by \cite{beam:Verkhodanov_n},
allow one to restore the
antenna
beam shape for pixelized beam on the $\Delta T$ map, which is different
from the antenna beam shape in the frame on the focal plane. In general
cases,
transition from a pixelized beam to the actual beam in the focal
plane frame requires a knowledge of the noise properties
\cite{beam:Verkhodanov_n}.

{\small
{\bf Acknowledgments.}
O.V.V. thanks the Russian Foundation for Basic Research for partial
support (grant no. 05-07-90139).
}

\end{document}